\documentclass[pra,twocolumn,smath,amssymb,superscriptaddress]{revtex4-1}

\usepackage{epsfig,amsmath}
\usepackage{subfigure}
\usepackage{graphicx}% Include figure files
\usepackage{dcolumn}% Align table columns on decimal point
\usepackage{stmaryrd}
\usepackage{mathrsfs}
\usepackage{pifont}
\usepackage{amsthm}
\usepackage{amssymb}
\usepackage{bm}
\usepackage{latexsym}
\usepackage[colorlinks=true,linkcolor=blue,citecolor=blue]{hyperref}
\usepackage{color}
\usepackage{epstopdf}

\newcommand{\non}{\nonumber}

\begin{document}

\title{High-capacity and high-power collective charging with spin chargers}

\author{Yong Huangfu}
\affiliation{Department of Physics, Zhejiang University, Hangzhou 310027, Zhejiang, China}

\author{Jun Jing}
\email{Email address: jingjun@zju.edu.cn}
\affiliation{Department of Physics, Zhejiang University, Hangzhou 310027, Zhejiang, China}

\date{\today}

\begin{abstract}
Quantum battery works as a micro- or nano-device to store and redistribute energy at the quantum level. Here we propose a spin-charger protocol, in which the battery cells are charged by a finite number of spins through a general Heisenberg XY interaction. Under the isotropic interaction, the spin-charger protocol is endowed with a higher capacity in terms of the maximum stored energy than the conventional protocols, where the battery is charged by a continuous-variable system, e.g., a cavity mode. By tuning the charger size, a trade-off between the maximum stored energy and the average charging power is found in comparison to the cavity-charger protocol in the Tavis-Cummings model. Quantum advantage of our protocol is manifested by the scaling behavior of the optimal average power with respect to the battery size, in comparing the collective charging scheme to its parallel counterpart. We also discuss the detrimental effect on the charging performance from the anisotropic interaction between the battery and the charger, the non-ideal initial states for both of them, and the crosstalk among the charger spins. A strong charger-charger interaction can be used to decouple the battery and the charger. Our findings about the advantages of the spin-charger protocol over the conventional cavity-charger protocols, including the high capacity of energy storage and the superior power-law in the collective charging, provide an insight to exploit an efficient quantum battery based on the spin-spin-environment model.
\end{abstract}

\maketitle

\section{Introduction}

Substantial effort in recent years has been devoted to promote the performance of the microscopic battery in the quantum regime, which plays an analog role of storing and extracting useful energy as its classical counterpart. Quantum battery is not only a scientifically interesting subject, but is a practical need. Devising a quantum battery in various forms has constituted an emerging field of nano-devices for quantum thermodynamics, such as solid-state batteries~\cite{Aykol-solidbattery-2019,Swift-solidbattery-2019}, topological batteries~\cite{Liu-topobattery-2018,Qie-topobattery-2018,Tserkovnyak-topobattery-2018,Patil-topobattery-2020}, and spin batteries~\cite{Brataas-spinbattery-2002,Xie-spinbattery-2018,Caravelli-spinbattery-2020}. The contemporary development of quantum battery also inspires the relevant investigations in quantum information about the work extraction~\cite{Frenzel-infromation-2014} and the optimal finite-time operation~\cite{Gallego-thermodynamics-2016,Anders-thermodynamics-2017,Sergi-Bounds-2020}.

The working procedure in quantum battery can be divided into three ordered steps: charging~\cite{Polini-HighPower-2018,Polini-mediated-2018,Crescente-Ultrafast-2020,Polini-SYK-2020,Polini-Manybody-2018,
Polini-Manybody-2019,Andolina-Manybody-2019,Chetcuti-Manybody-2020,Binder-global-2015,Kamin-ergotropy-2020,
Campaioli-parallel-2017, Sergi-Bounds-2020,Tabesh-Environment-2020}, energy storage~\cite{Gherardini-Stable-2020,Santos-Stable-2019,Santos-Stable-2020,Dvira-Stable-2019,An-Stable-2020}, and discharging~\cite{Allahverdyan-ergotropy-2004,Francica-ergotropy-2017,Alicki-battery-2013,Polini-ergotropy-2019,
Baris-ergotropy-2020}. The charger in convention is equipped by a continuous-variable system. The conventional charging protocol starts from the Jaynes-Cummings model~\cite{Polini-mediated-2018}, where the two-level system and the cavity mode serve as the battery cell and the charger system, respectively. The exact solution about the charging performance, including the maximum stored energy and optimal average charging power, provides a benchmark for the subsequent studies of the charging protocols assisted by a cavity mode. When the battery is extended to adopt a collection of two-level systems, one can have quantum batteries based on the Tavis-Cummings (TC) model~\cite{Polini-ergotropy-2019} and the Dicke model~\cite{Polini-HighPower-2018,Polini-mediated-2018,Crescente-Ultrafast-2020,Polini-SYK-2020}. When the charger becomes semi-classical, one can have many-spin batteries charged by a driving field~\cite{Polini-Manybody-2018,Polini-Manybody-2019,Andolina-Manybody-2019,Chetcuti-Manybody-2020}. Powerful charging protocols were reported by virtue of the two-photon process~\cite{Crescente-Ultrafast-2020}, the global operations~\cite{Binder-global-2015}, and the non-Markovian effects~\cite{Kamin-ergotropy-2020}. Many technologies were applied to a capable quantum battery for energy storage, such as the quantum measurement on Zeno protection~\cite{Gherardini-Stable-2020}, the stimulated Raman adiabatic passage technique~\cite{Santos-Stable-2019,Santos-Stable-2020}, the symmetry protected dark state living in a decoherence-free subspace~\cite{Dvira-Stable-2019}, and the Floquet engineering with two bound states in the quasienergy spectrum~\cite{An-Stable-2020}. To promote the work extraction, the amount of available energy from the active state to the passive state is studied in the discharging process~\cite{Francica-ergotropy-2017}. An observation is that the entangling unitary controls extract in general more work than the separable controls~\cite{Alicki-battery-2013}.

Among the studies of quantum battery in the collective scheme, quantum advantage is significantly reflected by the scaling behavior of the charging performance, especially of the charging power. The quantum spin battery in the Lipkin-Meshkov-Glick model~\cite{Sergi-Bounds-2020} shows that quantum entanglement enhances the charging power at the cost of the charging speed. In the Sachdev-Ye-Kitaev model~\cite{Polini-SYK-2020}, the power enhancement is linked to the variance of the charging Hamiltonian. In contrast to the parallel scheme in which each battery cell is individually charged, all cells in the collective scheme are charged through the global operations on the overall battery-charger system. The collective effect can boost the charging power of the quantum battery in terms of the scaling law of the number of battery cells $M$~\cite{Campaioli-parallel-2017}. The Dicke quantum battery gains a collective advantage with a $M^{1/2}$ scaling law~\cite{Polini-HighPower-2018}. It is soon followed by a strict comparison between the quantum battery and its classical counterpart~\cite{Andolina-Manybody-2019} to display the quantum collective advantage, in which the Dicke quantum battery holds a much better performance in the charging power.

In this work, we consider a closed quantum battery-charger system and focus only on the charging process via unitary, deterministic, and reversible evolutions. The discharging dynamics is then exactly the reverse process to the charging dynamics, i.e., the energy stored in the battery can be completely extracted in a reverse way. Rather than the conventional charging protocols assisted by a cavity mode, we study the charging performance in a protocol based on a spin-spin-environment model. In particular, the battery cells are coupled collectively to a finite number of non-interacting spin chargers through a general Heisenberg XY interaction. This model was significant for the central spins would hold coherence in a comparatively long time~\cite{Kurucz-cohtime-2009,Zadrozny-cohtime-2015,Abobeih-cohtime-2018} under a spin environment. It constitutes an important platform to investigate the entanglement dynamics~\cite{Xu-entanlement-2011,Yang-entanlement-2016,Jing-entanlement-2018} and the quantum control strategy~\cite{Wu-spincontrol-2010}. In this spin-charger protocol, it is interesting to find a trade-off between the maximum stored energy and the optimal average charging power by tuning the charger size and the collective advantage represented by the superior power-law scaling with respect to the battery size.

The rest of this work is organized as following. In Sec.~\ref{model}, we present the full Hamiltonian for the quantum spin batteries coupled to the charger spins. The charging performance is analysed under various anisotropic coefficient $\gamma$ and the ratios of the numbers of charger spins $N$ and battery cells $M$. All the results are compared with the conventional cavity-charging protocol based on TC model, that serves as a benchmark in the thermodynamical limit of our model. The dynamics of two spin batteries under the isotropic charging is provided in appendix~\ref{AppM2}, where the maximum stored energy is found to be consistent with the benchmark result. In Sec.~\ref{Qadvant}, we investigate the collective advantage of our spin-charger protocol against its parallel counterpart. In Sec.~\ref{nonideal}, we discuss the spin-charging protocol in the presence of either the charger-spin crosstalk or the non-ideal initial states. In Sec.~\ref{Con}, we summarize the whole work.

\section{Model of quantum battery under spin-chargers}\label{model}

\begin{figure}[htpb]
\centering
\includegraphics[width=0.45\textwidth]{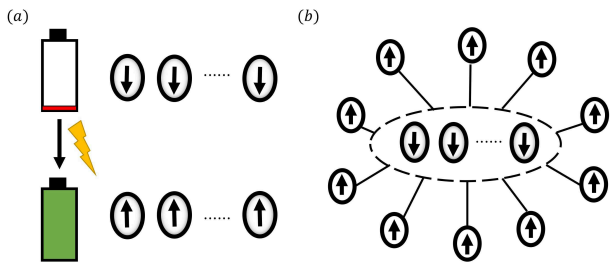}
\caption{(Color online) (a) The quantum battery is consisted of $M$ spin-$1/2$ cells. (b) The charging protocol is based on the spin-spin-environment model where $M$ quantum battery cells are coupled collectively to $N$ non-interacting charger spins.} \label{SpModel}
\end{figure}

In the quantum battery shown in Fig.~\ref{SpModel}, we consider $M$ spin-$1/2$ cells charged collectively by $N$ non-interacting charger spins-$1/2$. The overall Hamiltonian driving the charging process can be written as ($\hbar\equiv1$)
\begin{eqnarray}\label{Htol}
H &=& H_{0}+H_{1}, \\
H_{0} &=&\nonumber H_{\rm B}+H_{\rm C}=\frac{\omega_{0}}{2}\left(\sum^{M}_{k=1}\sigma^{z}_{k}+\sum^{N}_{l=1}\tau^{z}_{l}\right),  \\
H_{1} &=&\nonumber \frac{g}{2}\left[(1+\gamma)\sum^{M}_{k=1}\sigma^{x}_{k}\sum^{N}_{l=1}\tau^{x}_{l}
+(1-\gamma)\sum^{M}_{k=1}\sigma^{y}_{k}\sum^{N}_{l=1}\tau^{y}_{l}\right],
\end{eqnarray}
where $H_{\rm B}$, $H_{\rm C}$ and $H_1$ are the battery Hamiltonian, the charger Hamiltonian, and the battery-charger interaction Hamiltonian, respectively. The battery-spin and the charger-spin operators are denoted by the Pauli matrices $\sigma^{\alpha}_{k}$ and $\tau^{\alpha}_{l}$ with $\alpha\in\{x,y,z\}$, respectively. By convention, the battery cells are set to be resonant with the charger spins with frequency $\omega_0$. In the interaction Hamiltonian, $g$ is the coupling strength between batteries and chargers and $\gamma$, $0\leq\gamma\leq1$, is the anisotropic coefficient for coupling. When $\gamma=0$, the battery-charger interaction reduces to an isotropic XY interaction. When $\gamma=1$, it becomes an XX interaction with equal-weighted counter-rotating terms. The same model with the general Heisenberg XY interaction was applied to study the disentanglement dynamics of the central spins~\cite{Yuan-twospin-2007,Jing-HP-2007,Jing-twospin-2007}.

Using the collective angular momentum operators for the charger spins $J_{z,\pm}\equiv\sum^{N}_{l=1}\tau^{z,\pm}_{l}$ with $\tau^{\pm}_{k}\equiv(\tau^{x}_{k}\pm i\tau^{y}_{k})/2$, the charger Hamiltonian can thus be rewritten as
\begin{equation}\label{Hc}
  H_{\rm C}=\frac{\omega_{0}}{2}J_{z}.
\end{equation}
And the interaction Hamiltonian becomes
\begin{eqnarray}\nonumber
H_{1} &=& g\left[\sum^{M}_{k=1}\left(\gamma\sigma^{-}_{k}+\sigma^{+}_{k}\right)\sum^{N}_{l=1}\tau^{-}_{l}+{\rm H.c.}\right] \\
&=&\label{H1} g\left[J_{-}\sum^{M}_{k=1}\left(\gamma\sigma^{-}_{k}+\sigma^{+}_{k}\right)+{\rm H.c.}\right],
\end{eqnarray}
which is valid due to the indistinguishability of all of the charger spins. To study the charging performance, the quantum batteries and the spin chargers are prepared as the spin-down (empty or passive) and spin-up (full or active) states, respectively. Then the overall wavefunction reads,
\begin{equation}\label{ini}
|\Psi(0)\rangle=\prod^{M}_{k=1}|0_{k}\rangle\otimes\prod^{N}_{l=1}|1_{l}\rangle
=\prod^{M}_{k=1}|0_{k}\rangle\otimes|m=\frac{N}{2}\rangle.
\end{equation}
Note the eigenstates of the spin charger are denoted by $|m\rangle$'s, where $m$'s are the eigenvalues running from $-N/2$ to $N/2$ due to Eq.~(\ref{Hc}). $|m=N/2\rangle$ implies that the spin chargers are all in the excited state.

In the thermodynamical limit $N\rightarrow\infty$, the preceding Hamiltonian can be further simplified by the Holstein-Primakoff (HP) transformation~\cite{Holstein-HP-1940}:
\begin{equation}\label{HPtrans}
  J_{+}=b^{\dagger}\left(\sqrt{N-b^{\dagger}b}\right),\quad J_{-}=\left(\sqrt{N-b^{\dagger}b}\right)b.
\end{equation}
It is important to remark that here the bosonic operators $b^\dagger$ and $b$ do not directly correspond to the angular momentum operators $J_+$ and $J_-$, respectively. According to Eq.~(\ref{ini}), the initial state of the charger spins is a completely polarized state $|m=N/2\rangle$, i.e., the fully-up state, which is symmetric to another completely polarized state $|m=-N/2\rangle$, i.e., the fully-down state. To describe a high-spin system with a bosonic mode, it is often appropriate to use the HP transformation when the state of the high-spin system does not significantly deviate from a benchmark state, that could be either one of the completely polarized state.

Consequently, the interaction Hamiltonian in Eq.~(\ref{H1}) is revised to be a spin-boson Hamiltonian:
\begin{eqnarray}\nonumber
  H_{1}&=& g\left[\left(\sqrt{N-b^{\dagger}b}\right)b\sum^{M}_{k=1}\left(\gamma\sigma^{-}_{k}+\sigma^{+}_{k}\right)+{\rm H.c.}\right]\\ \label{H11} &=&\tilde{g}\left[b\sum^{M}_{k=1}\left(\gamma\sigma^{-}_{k}+\sigma^{+}_{k}\right)+{\rm H.c.}\right],
\end{eqnarray}
where the coupling strength $\tilde{g}\approx\sqrt{N}g$ in the thermodynamical limit. When $\gamma=0$, the transformed interaction Hamiltonian in Eq.~(\ref{H11}) describes the Tavis-Cummings (TC) model~\cite{Tavis-TCmodel-1968,Tavis-TCmodel-1969}. It is then used to provide a benchmark for our model in the limit of $N\rightarrow\infty$, while our spin-charger model focuses on the situation with a finite $N$ and most of our results are obtained by Eq.~(\ref{H1}). When $\gamma=1$, Eq.~(\ref{H11}) describes the Dicke model~\cite{Dicke-Dickemodel-1954} as a platform for superradiance. It can also be applied to a high-power charging protocol for quantum battery~\cite{Polini-HighPower-2018}.

With the initial state in Eq.~(\ref{ini}), the energy stored in the battery during the charging process is given by
\begin{equation}\label{SpEnergy}
  E(t)\equiv{\rm Tr}[\rho_{\rm B}(t)H_{\rm B}]-{\rm Tr}[\rho_{\rm B}(0)H_{\rm B}],
\end{equation}
where $\rho_{\rm B}(t)\equiv{\rm Tr}_{\rm C}[|\Psi(t)\rangle\langle\Psi(t)|]$. The average charging power is given by
\begin{equation}\label{SpPower}
 P(t)\equiv\frac{E(t)}{t}.
\end{equation}
The charging performance of a quantum battery can be measured respectively by $E_{\rm max}$ and $P_{\rm max}$ as the maximum value for both $E(t)$ and $P(t)$. In appendix~\ref{AppM2}, we provide the dynamics of both $E(t)$ and $P(t)$ as well as their optimized values for a special model in Eq.~(\ref{Htol}) with $M=2$ and $\gamma=0$.

\begin{figure}[htpb]
\centering
\includegraphics[width=0.45\textwidth]{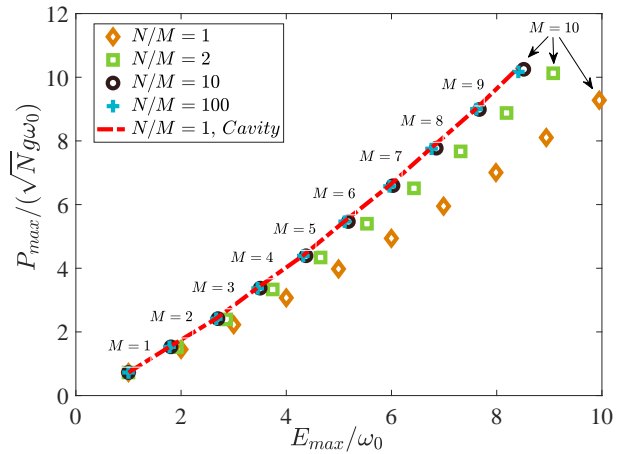}
\caption{(Color online) The landscape of the maximum stored energy $E_{\rm max}$ and the optimized average power $P_{\rm max}$ under the spin-charging protocol with isotropic interaction $\gamma=0$. Results of $N/M=1$, $N/M=2$, $N/M=10$, and $N/M=100$, are marked with orange diamonds, green squares, black circles and blue crosses, respectively. The red dot-dashed line indicates the benchmark result by the TC model, a conventional cavity-charging protocol. The battery-charger coupling strength is set as $g/\omega_{0}=0.1$.} \label{IsomaxEP}
\end{figure}

To distinguish the spin-charging protocol from the conventional cavity-charging protocol, we first present a landscape about the maximum stored energy $E_{\rm max}$ and the optimized average power $P_{\rm max}$ in Fig.~\ref{IsomaxEP} under the isotropic charging for various ratios of the charger-spin number $N$ and the battery-cell number $M$. The benchmark results of $E_{\rm max}$ and $P_{\rm max}$ in the TC model are displayed by the red dot-dashed line with $N/M=1$, where $N$ denotes the initial photon-number in the cavity mode. The average charging power $P_{\rm max}$ is normalized in the unit of $\sqrt{N}g\omega_0$ to be consistent with the rescaled coupling strength $\tilde{g}$ in the thermodynamical limit~\cite{Sergi-Bounds-2020}. It is found that the results in the spin-charging protocol are coincident with the benchmark results, when the ratio becomes sufficiently large $N/M\geq10$. This asymptotic behavior can be understood by virtue of the HP transformation, which maps the collective-spin ladder operators into the bosonic creation and annihilation operators. Then with a sufficiently large $N$, a spin charger becomes close to a cavity charger.

In Fig.~\ref{IsomaxEP}, our spin-charger protocol departs significantly from the TC model under a small ratio $N/M$. When $N/M=2$ (see the green squares), the maximum stored energy $E_{\rm max}$ becomes gradually larger than the corresponding TC result with an increasing $M$; while the optimized average powers $P_{\rm max}$ are nearly invariant. When $N/M=1$ (see the orange diamonds), a clear trade-off between $E_{\rm max}$ and $P_{\rm max}$ can be observed. In comparison to the TC model, a much higher stored energy can be achieved at the cost of a little lower optimized charing power when $M\geq3$. And a larger size of the spin-charger yields a larger $P_{\rm max}$ and a smaller $E_{\rm max}$ under a fixed battery size.

\begin{figure}[htpb]
\centering
\includegraphics[width=0.45\textwidth]{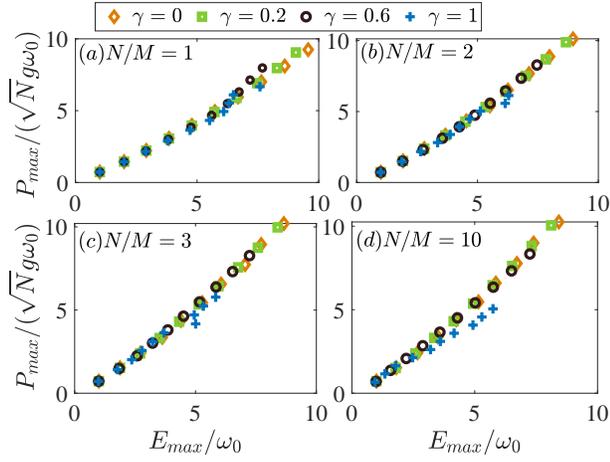}
\caption{(Color online) The landscape of $E_{\rm max}$ and $P_{\rm max}$ under the spin-charging protocol with various ratios of $N/M$: (a) $N/M=1$, (b) $N/M=2$, (c) $N/M=3$, (d) $N/M=10$. Results under the anisotropic coefficient $\gamma=0$, $0.2$, $0.6$, and $1$, are marked with orange diamonds, green squares, black circles and blue crosses, respectively. $g/\omega_{0}=0.1$. } \label{AnisomaxEP}
\end{figure}

Next we consider the impact of the anisotropic charging on the landscape of $E_{\rm max}$ and $P_{\rm max}$. In Fig.~\ref{AnisomaxEP}, again we collect the data for $M=1,2,\cdots,10$ with various ratios $N/M$. It is observed that all the data follow almost the same linear pattern irrespective to the value of $N/M$. Across all the four sub-figures, $E_{\rm max}$ is found to be proportional to $P_{\rm max}$ with the slope coefficients gradually enhanced from $1.002$ for $N/M=1$ to $1.307$ for $N/M=10$. A smaller anisotropic coefficient $\gamma$ yields a higher energy and a higher power. For example, in Fig.~\ref{AnisomaxEP}(a) with $M=10$, when $\gamma=1$, $E_{\rm max}/\omega_{0}\approx7.48$ and $P_{\rm max}/(\sqrt{N}g\omega_{0})\approx6.67$, and when $\gamma=0$, $E_{\rm max}/\omega_{0}\approx9.76$ and $P_{\rm max}/(\sqrt{N}g\omega_{0})\approx9.24$. The results of a strong anisotropic interaction for $\gamma=1$ will gradually deviate from the main line of $E_{\rm max}$ and $P_{\rm max}$ with an increasing $N/M$. In Fig.~\ref{AnisomaxEP}(d), it is clear that both $E_{\rm max}$ and $P_{\rm max}$ with $\gamma=1$ are considerably smaller than those with $\gamma=0$.

The anisotropic charging is associated mainly with the counter-rotating terms of the interaction Hamiltonian in Eq.~(\ref{H1}). During a unitary charging process, the energy is coherently transferred from the charger spins to the battery cells. Given the initial states in Eq.~(\ref{ini}), it is then preferable to promote the charging performance if the (empty) state of the battery can be exchanged with the (full) state of the charger. Note the counter-rotating terms cannot hold the energy conservation all the way. The charging dynamics is then mixed with the extra energy-generation or energy-annihilation process in the overall battery-charger system due to the non-commutativity between the interaction Hamiltonian $H_1$ in Eq.~(\ref{H1}) with a nonvanishing $\gamma$ and the free Hamiltonian $H_0$ in Eq.~(\ref{Htol}). The isotropic charging is therefore more favorable than the anisotropic charging. This result is in agreement with Ref.~\cite{Polini-mediated-2018} about the negative effects of the counter-rotating terms on the charging process.

\begin{figure}[htpb]
\centering
\includegraphics[width=0.45\textwidth]{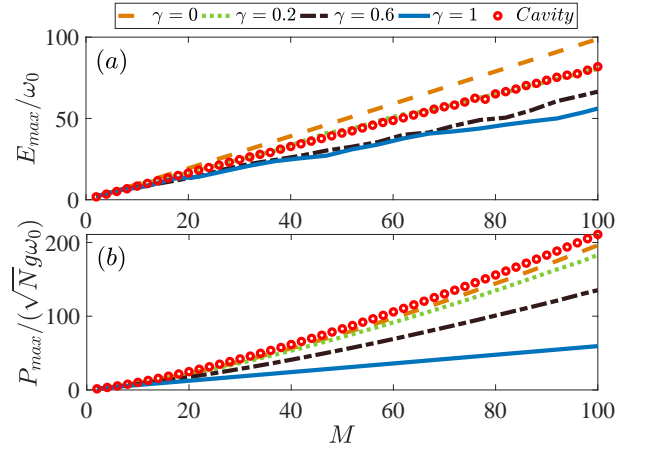}
\caption{(Color online) (a) $E_{\rm max}$ and (b) $P_{\rm max}$ as functions of the battery-cell number $M$ with various anisotropic coefficient. The results under $\gamma=0,0.2,0.6,1.0$ are plotted with the orange-dashed line, the green-dotted line, the black dot-dashed line, and the blue-solid line, respectively. The red circles indicate the benchmark result provided by the TC model. Here $N/M=1$ and $g/\omega_{0}=0.1$.} \label{M100maxEP}
\end{figure}

The dependence on the battery size of the charging performance, including the maximum stored energy $E_{\rm max}$ and the optimal average power $P_{\rm max}$, is shown in Fig.~\ref{M100maxEP}. For simplicity, here the battery and the charger are equivalent in their sizes, i.e., $N/M=1$. The cavity-charging protocol based on the TC model is again compared to the spin-charger protocol. As expected in Fig.~\ref{IsomaxEP} up to $M=10$, it is observed that the spin-charger protocol under the isotropic battery-charger interaction $\gamma=0$ surpasses the cavity-charging protocol in terms of $E_{\rm max}$, and more advantage can be distinguished by an even larger $M$. In contrast, the spin-charger protocol under $\gamma=0$ is slightly defeated by the benchmark protocol in terms of $P_{\rm max}$. In both Figs.~\ref{M100maxEP}(a) and (b), the anisotropic interaction degrades the charging performance, especially the charging power, in a monotonic way with increasing $\gamma$. In Fig.~\ref{M100maxEP}(a), it is found that both spin-charger and cavity-charger protocols attain almost the same $E_{\rm max}$ when $\gamma=0.2$.

\section{Quantum advantage of collective spin charging}\label{Qadvant}

Aside from the collective charging scheme studied in the previous section, an alternative option to implement the energy transfer for a many-cell battery is the parallel charging~\cite{Campaioli-parallel-2017}, by which each battery cell is separably charged by an individual charger. In our spin-charging model, the interaction Hamiltonian for the parallel scheme can be written as
\begin{equation}\label{H1-para}
H_{\parallel}=\sum^{M}_{k=1}h_k\equiv\sum^{M}_{k=1}g\left[J^{(k)}_{-}\left(\gamma\sigma^{-}_{k}+\sigma^{+}_{k}\right)
+{\rm H.c.}\right],
\end{equation}
where $h_k$ indicates the $k$th quantum cell in parallel. Both the overall maximum stored energy $E^{\parallel}_{\rm max}=Me_{\rm max}$ and the overall optimal average power $P^{\parallel}_{\rm max}=Mp_{\rm max}$ grow linearly with the number of battery cells $M$, where $e_{\rm max}$ and $p_{\rm max}$ represent the maximum stored energy and the optimal average power in a single battery cell, respectively. We can use the following ratio~\cite{Andolina-Manybody-2019}
\begin{equation}\label{eta}
  \eta\equiv\frac{P_{\rm max}}{P^{\parallel}_{\rm max}}
\end{equation}
to describe the quantum advantage of the collective charging scheme over the parallel counterpart, if $\eta>1$.

\begin{figure}[htpb]
\centering
\includegraphics[width=0.45\textwidth]{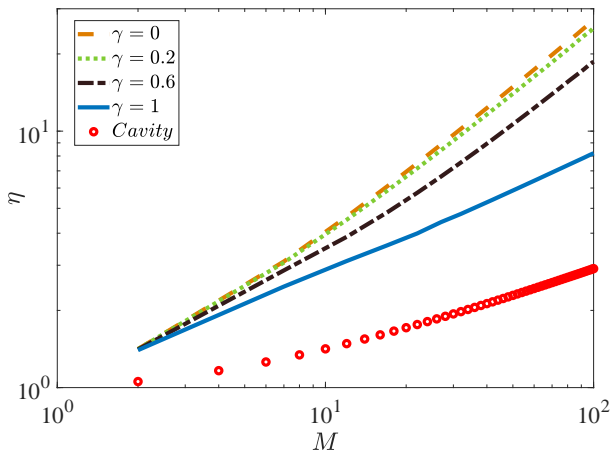}
\caption{(Color online) The ratio of the charging power under the collective-charging scheme and that under the parallel-charging scheme $\eta$ as a function of $M$. The results under $\gamma=0,0.2,0.6,1.0$ are plotted with the orange-dashed line, the green-dotted line, the black dot-dashed line and the blue-solid line, respectively. The red circles indicate the result of the TC model. $N/M=1$ and $g/\omega_{0}=0.1$. } \label{CollPall}
\end{figure}

The collective ratio $\eta$ as a function of $M$ under various anisotropic coefficients $\gamma$ are displayed in Fig.~\ref{CollPall} with $N/M=1$. It is noticeable that $\eta$ for the spin-charger protocol is remarkably larger than that for the cavity-charging protocol. In magnitude, the collective ratio $\eta$ roughly follows a power law of $M$:
\begin{equation}\label{QCAeta}
  \eta\propto M^{\beta}.
\end{equation}
The exponent index $\beta$ is found to gradually approach an asymptotic value by increasing $M$, which is near to unit when $\gamma=0$. In particular, when $M=20$, $\beta\approx0.81$; when $M=100$, $\beta\approx0.87$; and when $M=1000$, $\beta\approx0.99$. In Fig.~\ref{CollPall}, it is also observed that the index $\beta$ for $M=100$ decreases with an increasing $\gamma$. In particular, when $\gamma=0,0.2,0.6,1.0$, $\beta\approx0.87,0.84,0.79,0.48$, respectively. The result under $\gamma=0$ is consistent with a previous spin-charger protocol~\cite{Andolina-Manybody-2019} with a large $M$. Note it was found that $\beta=0.5$ for $M\gg1$ in the Dicke quantum battery~\cite{Polini-HighPower-2018} charged by a cavity mode, which is verified by the red circles plotted in Fig.~\ref{CollPall} under the TC model. Then our spin-charger protocol demonstrates a superior pow-law advantage than the cavity-charging protocols. The collective advantage could be understood by the effect from the global operations~\cite{Binder-global-2015}. In particular, the interaction between the array of quantum battery cells and the common chargers creates a quantum correlation giving rise to the advantage of the charging power over the parallel battery.

Additionally, the performance of the parallel charging does not depend on $\gamma$ as the collective charging when $N=M$. It is found to be completely determined by the single-cell result. Consider one cell charged by one spin-charger, which is ideally prepared as $|0_{\rm B}\rangle\otimes|1_{\rm C}\rangle$. In the space supported by $\{|1_{\rm B},1_{\rm C}\rangle$, $|1_{\rm B},0_{\rm C}\rangle$, $|0_{\rm B},1_{\rm C}\rangle$, and $|0_{\rm B},0_{\rm C}\rangle\}$, the total Hamiltonian can be written as
\begin{eqnarray}\nonumber
  h_1&=& \frac{\omega_{0}}{2}(\sigma_{z}+\tau_{z})+g[\tau_{-}(\gamma\sigma_{-}+\sigma_{+})+{\rm H.c.}] \\
  &=&\left(\begin{array}{cccc}
           \omega_{0} & 0 & 0 & g\gamma \\
           0 & 0 & g & 0 \\
           0 & g & 0 & 0 \\
           g\gamma & 0 & 0 & -\omega_{0} \\
         \end{array}
       \right).
\end{eqnarray}
The induced charging dynamics is localized in the subspace spanned by $|1_{\rm B},0_{\rm C}\rangle$ and $|0_{\rm B},1_{\rm C}\rangle$, irrelevant to the anisotropic coefficient $\gamma$. And it is straightforward to find that the stored energy $e(t)=\omega_{0}\sin^{2}(gt)$ and the average charging power $p(t)=\omega_{0}[\sin^{2}(gt)]/t$ for a single battery cell. Subsequently, the renormalized charging performance is given by $e_{\rm max}/\omega_{0}=1$ and $p_{\rm max}/g\omega_{0}=0.72$.

\section{Non-ideal charging}\label{nonideal}

Regarding the systematic errors, it is physically relevant to consider the charging performance with respect to the inner-coupling among spin chargers and the non-ideal initial states for both battery and charger. The former is associated with the situation where the charger spins are mutually coupled to each other with Heisenberg XY interaction, that would be inevitable due to the crosstalk among neighboring chargers. The latter might be induced by a finite fidelity in the state initialization of the battery or the charger. In this section, we only consider $\gamma=0$, i.e., the situation for the isotropic Heisenberg XY interaction. Then the interaction Hamiltonian reads,
\begin{equation}\label{H1int}
H_1=\frac{g}{2}\left(\sum^M_{k=1}\sigma^{x}_{k}\sum^N_{l=1}\tau^{x}_{l}
+\sum^M_{k=1}\sigma^{y}_{k}\sum^N_{l=1}\tau^{y}_{l}\right),
\end{equation}
which also serves as the main Hamiltonian in the rotating frame with respect to $H_0$. We choose a small size of quantum battery with $M=2$ and $N=4$, unless otherwise stated. One can compare directly the numerical results under these non-ideal situations with the analytical solution given in appendix~\ref{AppM2}. Also, it is numerically found that with a fixed set of parameters, these non-ideal impacts on our quantum battery with spin charger decline gradually with the system size. 

\subsection{The spin-charging protocol with inner-coupled chargers}

This subsection is contributed to the consequence from the crosstalk between charger spins. To simplify the discussion, the extra XY interactions among the charger spins are all the same in the coupling strength. A similar configuration for the spin-star network has been used to study the non-Markovian quantum dynamics of the central spins~\cite{Breuer-twospin-2004,Yuan-twospin-2007}. The Hamiltonian for the spin chargers in the rotating frame with respect to $H_0$ can then be written as
\begin{equation}\label{Hcint}
  H_{\rm C}=\frac{g_1}{2}\sum^{N}_{l\neq j}\left(\tau^{x}_{l}\tau^{x}_{j}+\tau^{y}_{l}\tau^{y}_{j}\right),
\end{equation}
where $g_{1}$ is the charger-charger coupling strength. Using the collective angular momentum operators, we have
\begin{equation}
  H_{\rm C}=g_1(J_{+}J_{-}+J_{-}J_{+}-\mathcal{I}),
\end{equation}
where $\mathcal{I}$ represents the identity operator, compensating the interaction terms with $l=j$ in Eq.~(\ref{Hcint}). Again the charging performance under the full Hamiltonian $H=H_1+H_{\rm C}$ is measured by the stored energy $E(t)$ the average charging power $P(t)$.

\begin{figure}[htpb]
\centering
\includegraphics[width=0.45\textwidth]{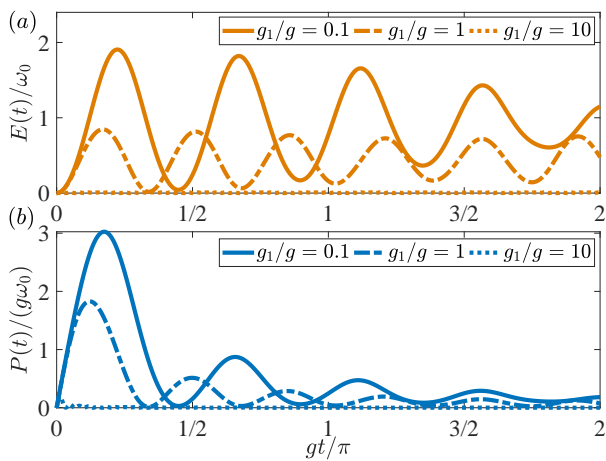}
\caption{(Color online) (a) The stored energy $E(t)$ and (b) the average charging power $P(t)$ against the dimensionless time $gt$ under various ratio of the charger-charger coupling strength and the battery-battery coupling strength $g_1/g$. Results for $g_{1}/g=0.1$, $g_{1}/g=1$, and $g_{1}/g=10$ are plotted with the solid, the dot-dashed, and the dotted lines, respectively.} \label{HcintM2N4}
\end{figure}

In Fig.~\ref{HcintM2N4}, we plot $E(t)$ and $P(t)$ with various $g_1/g$. The renormalized maximum stored energy is significantly reduced by increasing the charger-charger coupling. According to Eq.~(\ref{appEmax}), it is found that $E_{\rm max}/\omega_{0}=1.92$ with $M=2$, $N=4$ and $g_1/g=0$. When $g_{1}/g=0.1$, it is observed that $E_{\rm max}/\omega_{0}=1.91$ [see the solid line in Fig.~\ref{HcintM2N4}(a)]. When $g_{1}/g=1$ (see the dot-dashed line), $E_{\rm max}/\omega_{0}=0.85$. The numerical simulation shows that when $g_{1}/g\approx7$, $E_{\rm max}$ declines to about one percent of the energy for a full-charged state. And when $g_{1}/g=10$ (see the dotted line), the battery will completely lose its capacity to store energy. Similar pattern manifests in the dynamics of the average charging power $P(t)$. In contrast to the performance of an ideal battery with $M=2$ and $N=4$, i.e., $P_{\rm max}/(g\omega_{0})\approx3.04$, that can be numerically obtained by Eq.~(\ref{appEt}), a weaker charger-charger coupling $g_{1}/g=0.1$ renders almost the same result $P_{\rm max}/(g\omega_{0})\approx3.02$ [see the solid line in Fig.~\ref{HcintM2N4}(b)]; but a stronger coupling $g_{1}/g=1$ renders a remarkably lower result $P_{\rm max}/(g\omega_{0})\approx1.83$.

It is clear that the charger-charger interaction will damage the energy transfer from the full-charged spins to the empty-charged spins. The extra interaction of $H_{\rm C}$ in Eq.~(\ref{Hcint}) does not violate the energy conservation of the full battery-charger system. However, it will allow the energy to exchange among charger spins. That will affect the energy distribution of the whole system due to the comparative weights between $H_1$ in Eq.~(\ref{H1int}) and $H_{\rm C}$ in Eq.~(\ref{Hcint}). The deviation from the ideal situation can be estimated by the ratio $g_1/g$. As $g_1$ grows to overwhelm $g$, the battery cannot obtain any energy from the charger, but stays at the initial state [see the dotted lines in Figs.~\ref{HcintM2N4}(a) and (b)]. Thus the strong inner-coupling among charger spins will effectively decouple the battery and the charger.

One can also consider the crosstalk among the battery cells instead of that among the charger spins by exchanging these two spin-systems. Then the extra interaction of $H_{\rm C}$ in Eq.~(\ref{Hcint}) is replaced by $H_{\rm B}=(g_2/2)\sum_{k\neq m}^M(\sigma_{k}^x\sigma_{m}^x+\sigma_{k}^y\sigma_{m}^y)\simeq g_2(J_+J_-+J_-J_+)$, where $g_2$ denotes the inner-coupling strength among battery cells and $J_{\pm}$ is now understood as $\sum_{k=1}^M\sigma_k^{\pm}$. Then through a parallel analysis as in Fig.~\ref{HcintM2N4}, one can anticipate that the charge performance will decline in the presence of a weak $g_2$ due to the energy redistribution in the quantum battery. A sufficiently strong $g_2$ will have the same effect as a dominant $g_1$ by effectively separating the two spin-systems, since now $H_1$ becomes ignorable. In addition, the inner-coupling energy by $H_{\rm B}$ will be transferred to the quantum battery, given the initial empty-state for the battery cells. This ``self-charging'' phenomenon is however beyond our consideration.

\subsection{The spin-charging protocol with non-ideal initial states}

The spin-charging protocol starts with the full-charged chargers and the empty-charged batteries in the ideal situation. In this subsection, we discuss the impacts from the charger prepared at the partially-charged state and the battery with partially-charged cells, respectively.

It is heuristic to first consider one of the charger spins in a thermal-equilibrium state. Take a model in which two cells are charged by two charger spins as an example. Assume the $2$nd charger spin is in a thermal state as
\begin{equation}\label{inicM2N2}
  \rho^{(2)}_{\rm C}(t=0)=\left(
                        \begin{array}{cc}
                         p_1& 0 \\
                          0 & p_0 \\
                        \end{array}
                      \right),
\end{equation}
where $p_1$ and $p_0$ determine the effective (positive or negative) temperature and $p_1+p_0=1$. Thus the initial state of the whole system can be written as $\rho_{\rm BC}(0)=|00\rangle\langle00|\otimes(p_1|11\rangle\langle11|+p_0|10\rangle\langle10|)$. The time evolution of the stored energy for the battery is given by
\begin{eqnarray}\non
  E(t)&=&2\omega_0\sin^2\left(\sqrt{2}gt\right) \\
  &&-p_0\omega_0\left[2\sin^2\left(\sqrt{2}gt\right)-\sin^2(2gt)\right].
\end{eqnarray}
It reduced to Eq.~(\ref{appEt}) when $p_0=0$. Clearly, the full charging cannot be realized in the presence of a nonvanishing $p_0$.

Generally the initial-state preparation for both battery and charger is a challenge in experiments~\cite{Yang-entanlement-2016}. Regarding the non-ideal spin chargers, one can write the wavefunction of the overall system as
\begin{equation}
  |\Psi^{\rm C}_{l}\rangle=\prod_{k}|0_{k}\rangle\otimes|m=-\frac{N}{2}+l-1\rangle,
\end{equation}
which means $l-1$ charger spins are initially at their excited level, $1\leq l\leq N$. Thus in general, the overall density matrix can be written as
\begin{equation}\label{nideini}
  \rho_{\rm BC}=p_{0}|\Psi(0)\rangle\langle\Psi(0)|+\sum^{N}_{l=1}p_{l}|\Psi^{\rm C}_{l}\rangle\langle\Psi^{C}_{l}|,
\end{equation}
where $|\Psi(0)\rangle$ is the ideal initial state given in Eq.~(\ref{ini}) and $\sum^{N}_{l=0}p_{l}=1$, $p_l\geq0$. Note Eq.~(\ref{nideini}) can be used to roughly estimate the effect from the charger spins in a mixed state, such as the thermal state in Eq.~(\ref{inicM2N2}).

Regarding the non-ideal quantum battery, where the $k$th cell is prepared at the excited state, the wavefunction of the overall system can be written as
\begin{equation}
  |\Psi^{\rm B}_{k}\rangle=\sigma^{x}_{k}\prod_{q}|0_{q}\rangle\otimes|m=\frac{N}{2}\rangle,
\end{equation}
and then the initial density matrix is in general given by
\begin{equation}\label{nempini}
  \rho_{\rm BC}=\tilde{p}_{0}|\Psi(0)\rangle\langle\Psi(0)|+\sum^{M}_{k=1}\tilde{p}_{k}|\Psi^{\rm B}_{k}\rangle\langle \Psi^{B}_{k}|,
\end{equation}
where $\sum^{M}_{k=0}\tilde{p}_{k}=1$ with $\tilde{p}_{k}\geq0$.

\begin{figure}[htpb]
\centering
\includegraphics[width=0.45\textwidth]{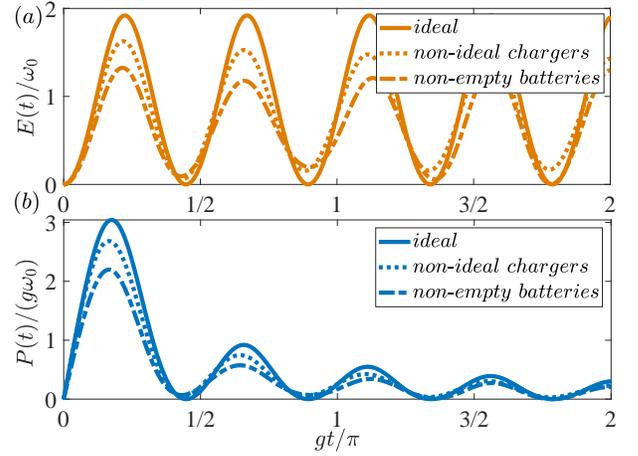}
\caption{(Color online) (a) The stored energy $E(t)$ and (b) the average charging power $P(t)$ against the dimensionless time $gt$ under different initial states. The probability distribution for the initial state of the non-ideal chargers and the non-empty batteries are set as $p_{0}=0.6$, $p_{1}=p_{2}=p_{3}=p_{4}=0.1$ and $\tilde{p}_{0}=0.6$, $\tilde{p}_{1}=\tilde{p}_{2}=0.2$, respectively.} \label{nideEP}
\end{figure}

The charging dynamics for the stored energy $E(t)$ and the average charging power $P(t)$ are plotted in Figs.~\ref{nideEP}(a) and (b), respectively with $p_{0}=0.6$, $p_{1}=p_{2}=p_{3}=p_{4}=0.1$ for the non-ideal chargers and $\tilde{p}_{0}=0.6$, $\tilde{p}_{1}=\tilde{p}_{2}=0.2$ for the non-empty batteries. In contrast to the ideal situation (see the solid lines), the non-ideal initial states of the charger impede the full-charging process, due to the existence of $|\Psi^{\rm C}_{1}\rangle=|00,m=-2\rangle$ and $|\Psi^{\rm C}_{2}\rangle=|00,m=-1\rangle$. Under the chosen probability distribution in Fig.~\ref{nideEP}, the optimal stored energy and the charging power decline respectively to about $84\%$ and $88\%$ of the corresponding ideal results. On the other hand, the reduction of the charging performance under the non-empty battery is caused by the quantum interference between the identical states $|\Psi^{\rm B}_{1}\rangle=|01,m=2\rangle$ and $|\Psi^{\rm B}_{2}\rangle=|10,m=2\rangle$. The charging process from the empty state $|00\rangle$ is then not synchronous with that from the non-empty state $|01\rangle$ or $|10\rangle$. Comparing to the non-ideal chargers (see the dotted lines), the non-empty batteries (see the dot-dashed lines) give rise to an even worse behavior in terms of charging performance. The maximum stored energy and the optimal average charging power are about $70\%$ and $72\%$, respectively, of the results in the ideal situation.

These results could also be theoretically anticipated since the non-ideal states are passive states from which no work can be extracted~\cite{Farina2019}. The mixture of the non-ideal states and the ideal states affects detrimentally to the charging performance.

\section{Conclusion}\label{Con}

In this work, we investigate the charging capacity and power of a quantum battery in a spin-charger protocol, which is established in a spin-spin-environment model through the general Heisenberg XY interaction. In comparison to the conventional cavity-charger protocol, the charging performance of our spin-charger protocol is examined with respect to its dependence on the size of both battery and charger and the anisotropic coefficient of the battery-charger coupling. Exploiting the collective coupling between battery cells and charger spins, it is interesting to find that: (1) a full charging can be achieved when the charger spins are coupled to the same number of the battery cells, that is beyond the reach of the conventional cavity-charging protocols; (2) the influences from the finite spin-charger size on the maximum stored energy and the optimized average charging power of quantum battery are opposite to each other, i.e., a larger N gives rise to a higher $P_{max}$ but a lower $E_{max}$; (3) the anisotropic battery-charger coupling will cut down the charging performance. We also investigate the quantum advantage of the collective-charging scheme over the parallel-charging scheme. It can be demonstrated by a superior power law than the cavity-charging protocol with respect to the battery size. Checking the performance of the quantum battery under spin-charging in the non-ideal situations, it is found that a sufficiently strong inner-coupling between the charger spins can be used to separate the battery and charger systems. Our spin-charging protocol with the advantage of high-capacity and high-power collective charging can be practiced in a solid-state platform and sheds new light on the intersection between the spin-spin-environment model and quantum thermodynamics.

\section*{Acknowledgements}

We acknowledge grant support from the National Science Foundation of China (Grants No. 11974311 and No. U1801661), and Zhejiang Provincial Natural Science Foundation of China (Grant No. LD18A040001).

\appendix

\section{Two-spin battery coupled to $N$-spin charger}\label{AppM2}

In this appendix, we concentrate on the charging performance of the battery system consisted of two spins coupled to $N\geq M=2$ charger spins in an isotropic Heisenberg XY model, i.e., $\gamma=0$. The total Hamiltonian in Eq.~(\ref{Htol}) becomes
\begin{eqnarray}\label{appHtolM2}
  H &=& H_{0}+H_{1}, \\
  H_{0} &=&\nonumber H_{\rm B}+H_{\rm C}=\frac{\omega_{0}}{2}(\sigma^{z}_{1}+\sigma^{z}_{2})+\frac{\omega_{0}}{2}J_{z},  \\
  H_{1} &=&\nonumber g\left[J_{-}(\sigma^{+}_{1}+\sigma^{+}_{2})+J_+(\sigma^-_{1}+\sigma^-_{2})\right],
\end{eqnarray}
which is equivalent to performing a rotating-wave approximation to the interaction Hamiltonian with general Heisenberg XY interaction. The model could be solved in the rotating frame with respect to $H_0$. Given the initial state in Eq.~(\ref{ini}), the Hamiltonian could be expressed in the subspace spanned by $|\lambda_{1}\rangle=|11,N/2-2\rangle$, $|\lambda_{2}\rangle=|10,N/2-1\rangle$, $|\lambda_{3}\rangle=|01,N/2-1\rangle$, and $|\lambda_{4}\rangle=|00,N/2\rangle$ as
\begin{equation}\label{appHSM2}
  H_{I}=g\left(
      \begin{array}{cccc}
        0 & \sqrt{2N-2}  & \sqrt{2N-2} & 0 \\
        \sqrt{2N-2} & 0 & 0 & \sqrt{N} \\
        \sqrt{2N-2} & 0 & 0 & \sqrt{N} \\
        0 & \sqrt{N}& \sqrt{N} & 0 \\
      \end{array}
    \right).
\end{equation}
In contrast, the Hamiltonian for the cavity-charging protocol in the TC model, where the battery spins are charged by a single-mode cavity, is written as
\begin{equation}\label{appHcavity}
  \tilde{H}=\omega_{0}b^{\dagger}b + \frac{\omega_{0}}{2}\sum^{M}_{k=1}\sigma^{z}_{k} + \tilde{g}\left(b\sum^{M}_{k=1}\sigma^{+}_{k}+ {\rm H.c.}\right).
\end{equation}
Here $b$ ($b^{\dagger}$) denotes the annihilation (creation) operator for the resonant mode and $\tilde{g}$ is the coupling strength between the quantum spin batteries and the mode $b$. Given the initial state $|00\rangle\otimes|2\rangle_b$, the rotating Hamiltonian with respect to $\omega_{0}b^{\dagger}b+(\omega_{0}/2)\sum_k\sigma^{z}_{k}$ could be written as
\begin{equation}\label{appHcavityM2}
  \tilde{H}_{I}=\tilde{g}\left(
      \begin{array}{cccc}
        0 & \sqrt{2}  & \sqrt{2} & 0 \\
        \sqrt{2} & 0 & 0 & 1 \\
        \sqrt{2} & 0 & 0 & 1 \\
        0 & 1& 1 & 0 \\
      \end{array}
    \right),
\end{equation}
where the bases are ordered by $|00,2\rangle$, $|01,1\rangle$, $|10,1\rangle$, and $|11,0\rangle$. In the thermodynamical limit with $N\rightarrow\infty$~\cite{Sergi-Bounds-2020,Andolina-Manybody-2019}, the two Hamiltonian in Eqs.~(\ref{appHSM2}) and (\ref{appHcavityM2}) are equivalent to each other by virtue of the Holstein-Primakoff transformation under the setting $\tilde{g}\approx\sqrt{N}g$.

Starting from the state $|\Psi(0)\rangle=|\lambda_{4}\rangle=|00,N/2\rangle$, the time evolution of the whole system driven by the Hamiltonian in Eq.~(\ref{appHSM2}) is
\begin{equation}
|\Psi(t)\rangle=e^{-iH_It}|\Psi(0)\rangle=\sum^{4}_{k=1}c_{k}(t)|\lambda_{k}\rangle,
\end{equation}
where
\begin{eqnarray*}
  c_{1}(t) &=& -\frac{2\sqrt{2(\xi^2+4)(\xi^2-2)}}{3\xi^2}\sin^{2}\left(\frac{\xi gt}{2}\right), \\
  c_{2}(t) &=& c_{3}(t)= -i\sqrt{\frac{\xi^2+4}{6\xi^2}}\sin(\xi gt), \\
  c_{4}(t) &=& \frac{1}{3\xi^2}\left[(\xi^2+4)\cos(\xi gt)+2(\xi^2-2)\right].
\end{eqnarray*}
with $\xi\equiv\sqrt{2(3N-2)}$.

\begin{figure}[htpb]
\centering
\includegraphics[width=0.45\textwidth]{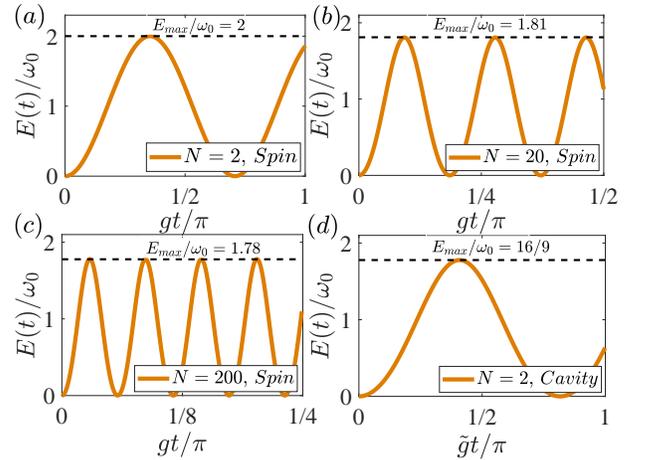}
\caption{(Color online) The stored energy $E(t)$ of the two-battery-cell system against the dimensionless time $gt$ with various numbers of charger spins: (a) $N=2$ with $E_{\rm max}/\omega_{0}=2$, (b) $N=20$ with $E_{\rm max}/\omega_{0}=1.81$, (c) $N=200$ with $E_{\rm max}/\omega_{0}=1.78$. (d) presents the benchmark result obtained in the TC model with $N=2$ and $E_{\rm max}/\omega_{0}=16/9$.} \label{EtM2}
\end{figure}

\begin{figure}[htbp]
\centering
\includegraphics[width=0.45\textwidth]{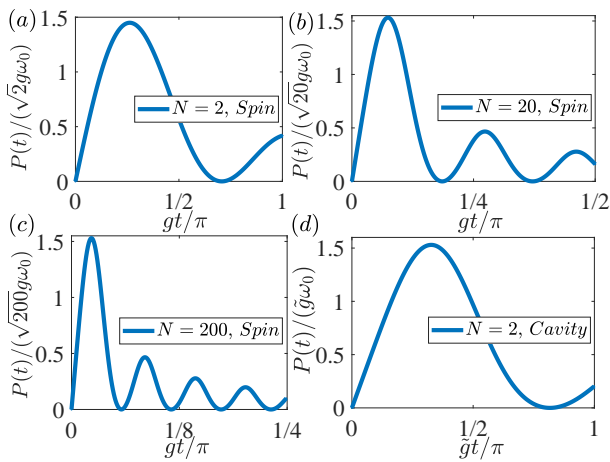}
\caption{(Color online) The average charging power $P(t)$ of the two-battery-cell system against the dimensionless time $gt$ with various numbers of charger spins: (a) $N=2$, (b) $N=20$, (c) $N=200$. (d) presents the benchmark result obtained in the TC model with $N=2$.} \label{PtM2}
\end{figure}

According to Eq.~(\ref{SpEnergy}), the time evolution of the energy stored in the double-spin quantum battery is given by
\begin{eqnarray}
  E(t) &=&\nonumber  \omega_{0}\left[|c_{1}(t)|^2-|c_{4}(t)|^2\right]+\omega_{0}\\
   &=&\nonumber\frac{(\xi^2+4)\omega_{0}}{9\xi^4}\left[(\xi^2-8)\cos(\xi gt)-(7\xi^2-8)\right]\\ \label{appEt}
       &\times&\left[\cos(\xi gt)-1\right].
\end{eqnarray}
Then it is straightforward to find the maximum stored energy measuring the capacity of the quantum battery,
\begin{equation}\label{appEmax}
 E_{\rm max}=\frac{16(\xi^4+2\xi^2-8)}{9\xi^4}\omega_{0}=\frac{16N(N-1)}{(3N-2)^2}\omega_{0},
\end{equation}
at $t=\bar{t}=\pi/(\xi g)$ for the first time. When $N\gg1$, $E_{\rm max}\rightarrow(16/9)\omega_0$. This result is consistent with that obtained by the cavity-charging protocol in the TC model~\cite{Polini-HighPower-2018}. We plot the stored energy $E(t)$ with various $N$ in both spin-charging protocol and the cavity-charging protocol in Fig.~\ref{EtM2}. As found in Eqs.~(\ref{appEt}) and (\ref{appEmax}), it is observed that $E(t)$ shows an ever faster sinusoid evolution with an increasing $N$, and then the optimized time $\bar{t}$ when $E(t)$ attains the maximum value $E_{\rm max}$ for the first time is consequently reduced. More importantly, $E_{\rm max}/\omega_0$ gradually approaches $16/9$, the value for the cavity-charging protocol as depicted in Fig.~\ref{EtM2}(d). The spin-charging protocol with a finite size thus activates a higher-capacity quantum battery in terms of both $E(t)$ and $E_{\rm max}$.

The average charging power $P(t)$ can be obtained through numerical simulation according to Eq.~(\ref{SpPower}). In all the sub-figures in Fig.~\ref{PtM2}, it is shown that in general after the power $P(t)$ increases to a maximum value $P_{\rm max}$, that could be conveniently obtained by an optimization procedure. And then it will asymptotically decay with time. This pattern is irrespective to the charging protocols, although a larger number of spin-charger $N$ brings more fluctuations during the evolution. The renormalized optimal charging-power is found to be $P_{\rm max}/(\sqrt{N}g\omega_0)\approx1.45$ when $N=2$ [see Fig.~\ref{PtM2}(a)], and it is slightly enhanced to $1.53$ when $N=200$ [see Fig.~\ref{PtM2}(c)]. The latter is close to the result obtained in the cavity-charging protocol as shown in Fig.~\ref{PtM2}(d).

Combining Figs.~\ref{EtM2} and \ref{PtM2}, a trade-off relation with respect to the charger size $N$ thus emerges between the maximum stored energy and the optimal average power.

\bibliographystyle{apsrevlong}
\bibliography{reference}

\end{document}